# Implementation of analytic functions with quantum gates

G. Florio[a] and D. Picca

Dipartimento Interateneo di Fisica "M. Merlin", Università di Bari, and INFN (Sezione di Bari)

Via Orabona 4, 70126, Bari, Italy

Abstract

In order to realize a "Quantum CPU" some schemes for executing fundamental mathematical tasks are needed. In this paper we present some quantum circuits which, using elementary arithmetic operations, allow an approximated calculation of continuous functions. Furthermore, we give an explicit example of our procedure applied to the exponential function.

**Introduction**

In recent years quantum computation has received great attention because of its possibility to solve important problems in number theory (factoring large integers), security (quantum cryptography) and searching in database (Grover's algorithm). Despite of the fact that experimental implementations are still "work in progress" (there are several devices and techniques [1]) and all present unsolved problems such as decoherence and read-out procedures, physicists keep trying to conceive possible schemes to implement known algorithms [2,3,4,5].

In this work we briefly show some simple quantum schemes for elementary arithmetic operations and then we use them to build more complex circuits. We focus our attention more on simplicity of execution than on efficiency of the procedure: this is clearly done for experimental purpose. In principle, this architecture can be used for calculating the values of a generic function. Here we limit to the simple exponential $e^x$.

---

[a] E-mail address: giuseppe.florio@ba.infn.it

# 1 - Quantum Adder

The first circuit capable of executing sum was proposed by Vedral et al. [5]; it makes use of some supplementary memory registers to store information about the carry.

As observed by Draper [6], the use of the Quantum Fourier Transform (qFT) allows the calculus to be executed saving memory. In more detail, let us consider two integer numbers **a** and **b** stored in distinct registers, each constituted by n qubits; if the circuit for qFT is applied to the ket $|a\rangle = |a_1 a_2 ... a_n\rangle$, the representative state of number **a**, its most significant qubit will be in a superposition of the form

$$|\varphi_1(a)\rangle = \frac{(|0\rangle + e^{2\pi i 0.a_1 a_2 ... a_n}|1\rangle)}{2^{1/2}} \qquad (1.1)$$

where the expression $0.a_1 a_2..a_n$ represents a binary fraction; the other qubits will be in similar superposition with different binary fraction.

At this point, we apply the circuit shown in fig. 1.1 where we only use quantum phase gate $R_k$ with the form

$$R_k = \begin{pmatrix} 1 & 0 \\ 0 & e^{2\pi i/2^k} \end{pmatrix} \qquad (1.2)$$

For explaining clearly this procedure, we trace the state of the most significant qubit during the calculation:

$$|\varphi_1(a)\rangle \rightarrow \frac{1}{2^{1/2}}(|0\rangle + e^{2\pi i 0.a_1 a_2 ... a_n + 0.b_1}|1\rangle) \qquad \text{after } R_1 \text{ from } b_1$$

$$\rightarrow \frac{1}{2^{1/2}}(|0\rangle + e^{2\pi i 0.a_1 a_2 ... a_n + 0.b_1 b_2}|1\rangle) \qquad \text{after } R_2 \text{ from } b_2$$

$$\ldots\ldots\ldots\ldots\ldots\ldots\ldots\ldots\ldots$$

$$\rightarrow \frac{1}{2^{1/2}}(|0\rangle + e^{2\pi i 0.a_1 a_2 ... a_n + 0.b_1 b_2 ... b_n}|1\rangle) \qquad \text{after } R_n \text{ from } b_n$$

$$= |\varphi_1(a+b)\rangle$$

After applying this procedure to all qubits, the circuit for inverse qFT can be used to obtain the representative state of the number **a+b**. We note that there is no need for supplementary register to store the carry qubits.

The circuit of fig. 1.1 is very similar to the qFT; the number of gates needed to accomplish the operation can be estimated [6] in **nlog$_2$n** (where n is the amount of qubits used); the main difference between qFT and quantum addition is the absence of Hadamard gates in the latter. As a consequence, all the operations in commute and thus they can be applied in arbitrary sequence. In fig. 1.2, the scheme of the modified circuit is shown. We immediately note that all $R_1$ can be performed at the same time because they involve distinct qubits. This applies to $R_2$ and so on; as a result, the number of operations is the same but the time needed for the calculus is reduced by **n** times and so it is O(log$_2$n).

**2 - Quantum Multiplier**

In this section, a circuit for multiplication will be explicitly derived.

This arithmetic operation can be considered as a kind of "extended sum". Let us consider two integer numbers **a** and **b**; the former is called "multiplicand", the latter is the "multiplier" as factors of the product. The multiplication algorithm basically consists in repeated sum of **a** for **b** times; the flowchart in fig. 2.1 shows this procedure explicitly.

The value of the multiplier is stored in a register labeled D, while register A is initially empty (i.e. it is in a fundamental state, for instance 0); first, the multiplicand is memorized in A and D is decremented by 1; if a fundamental value (0) is found in D, the calculus can be stopped; otherwise, the multiplicand is again added to A. This procedure is repeated until register D is empty (i.e. in the fundamental state).

The same scheme can also be used in the quantum domain.

The circuit in fig. 2.2 represents the general procedure for executing a multiplication. Let us consider two numbers **x** and **y** stored in the appropriate registers; let us suppose that both can be coded in n qubits; These numbers can now be considered as vectors in the respective computational bases and labeled by $|x\rangle$ and $|y\rangle$. The register marked with "Control" is composed by one qubit and is required for stopping the procedure. The register "Accumulator" must be composed by a number of qubits sufficient for storing the product: indeed, it is here that the operation results are progressively obtained. The controlled gate D is applied when the control register is in the state $|0\rangle$ and decreases by 1 the value of the number coded in the state $|y\rangle$; its scheme is shown in fig. 2.3. Its action, if enabled, is analogous to that of the circuit for quantum sum; qFT and qFT$^{-1}$ represent respectively direct and inverse quantum Fourier transform; it is important to remark that in this case we use gate (-R$_k$) in which the phase is negative; an explicit matrix representation is

$$-R_k = \begin{pmatrix} 1 & 0 \\ 0 & e^{-\frac{2\pi i}{2^k}} \end{pmatrix} \qquad (2.1)$$

An example will clarify this procedure. Let us consider the integer number 3: in a binary form it is expressed by 11 so it can be coded in state $|11\rangle$; we want to subtract integer 1 (coded in $|01\rangle$):

$$|11\rangle \xrightarrow{qFT} \begin{cases} \frac{1}{\sqrt{2}}(|0\rangle + e^{2\pi i(\frac{1}{2}+\frac{1}{4})}|1\rangle) \\ \frac{1}{\sqrt{2}}(|0\rangle + e^{2\pi i(\frac{1}{2})}|1\rangle) \end{cases} \xrightarrow{R} \begin{cases} \frac{1}{\sqrt{2}}(|0\rangle + e^{2\pi i(\frac{1}{2}+\frac{1}{4}-\frac{1}{4})}|1\rangle) \\ \frac{1}{\sqrt{2}}(|0\rangle + e^{2\pi i(\frac{1}{2}+\frac{1}{2})}|1\rangle) \end{cases} =$$

$$= \begin{cases} \frac{1}{\sqrt{2}}(|0\rangle + e^{2\pi i(\frac{1}{2})}|1\rangle) \\ \frac{1}{\sqrt{2}}(|0\rangle + e^{2\pi i(1)}|1\rangle) \end{cases} \xrightarrow{qFT^{-1}} |10\rangle$$

We have obtained state $|10\rangle$ that represents integer 2. this procedure can be repeated until we obtain state $|00\rangle$ obviously representing 0. Thus we can set to zero the multiplier register.

Another circuit block is the doubly controlled gate "S". Its action is that of summing the multiplicand to the contents of the accumulator by using the quantum adder.

C-NOT gates after D gates control if the contents in the $|y\rangle$ register are $|00....0\rangle$; in this case, the control qubit passes from state $|0\rangle$ to state $|1\rangle$ disabling all the next D and S gates and stopping the multiplication; actually the D and S gates are not two-qubits gates; their activation is controlled by all the qubits in the $|y\rangle$ register.

Let us trace the procedure step by step.

Accumulator and control registers are loaded respectively with states $|00....0\rangle$ (the number of qubit depends on the resources needed to store the product) and $|0\rangle$; $|x\rangle$ and $|y\rangle$ registers are used to store the multiplicand and multiplier. qFT is applied to the accumulator to prepare the system to repeated sums; the first C-NOT gate is needed to check $|y\rangle$ and verify if it is in the fundamental state $|00....0\rangle$; if not, Controlled-S gate is enabled and $|x\rangle$ is stored in the accumulator in the Fourier form; precisely we have

$$\frac{(|0\rangle + e^{2\pi i 0.x_1 x_2 ... x_n}|1\rangle)......(|0\rangle + e^{2\pi i 0.x_{n-1} x_n}|1\rangle)(|0\rangle + e^{2\pi i 0.x_n}|1\rangle)}{2^{n/2}} \qquad (2.2)$$

The state $|y\rangle$ is transformed into $|y-1\rangle$ by the D gate and another C-NOT check the state in the register; if it is still different from the fundamental, the procedure is repeated summing again $|x\rangle$ to the accumulator; we obtain (for simplicity we show only the most significant qubit):

$$\frac{1}{\sqrt{2}}(|0\rangle + e^{2\pi i (x_1/2 + x_2/4 + ... + x_n/2^n + x_1/2 + x_2/4 + ... + x_n/2^n)}|1\rangle) \qquad (2.3)$$

which represents the most significant qubit of the state $|2x\rangle$.

The procedure will continue until the $|y\rangle$ register reaches zero; C-NOT is then enabled and therefore the control qubit is commuted from $|0\rangle$ to $|1\rangle$; next S and D gates are disen-abled; last gate (qFT$^{-1}$) is applied to the accumulator register extracting the state $|x \bullet y\rangle$. As a result, the product is obtained.

## 3 - The general technique

In general, the technique used in modern computers for calculating the value of a non-elementary function f(x) consists essentially in approximating it in successive steps by a development in power series. A function of class C$^n$ (i.e. whose derivatives exist and are continuous until order n) in an interval (a,b), for every $x \in (a,b)$ can be expanded as:

$$f(x) = \sum_{k=0}^{n} \frac{f^{(k)}(x_0)}{k!}(x-x_0)^k + O[x-x_0]^{n+1} \tag{3.1}$$

Circuit implementation is based on truncating the series to the n-th power. By this way, it is possible to reduce the calculus to a sequence of additions and multiplications. It is obvious that more terms are in (3.1) higher the accuracy in the determination of f(x). The only obvious limitations are the calculation time and the memory resources of the computer.

## 4 – The circuit for the approximate calculus

In this paragraph we consider a specific case, the exponential function although the same technique can be used for a more general continuous function.

The function $f(x) = e^x$ can be developed in power series in R (real numbers field) and can be written (with initial point $x_0 = 0$) as:

$$e^x = \sum_{k=0}^{\infty} \frac{1}{k!} x^k = 1 + x + \frac{1}{2}x^2 + \frac{1}{6}x^3 + .... \tag{4.1}$$

The development consists in a sum of terms with form x$^k$ "weighted" by factors 1/k!. Then we can build the circuit with several "modules", each of them executing a particular procedure. Fig. 4.1 shows the proper scheme.

We consider a module that calculate all power terms x$^k$; each of them is multiplied with the relative weight 1/k! and the result is stored in a register where, by progressive summation, the approximate value of e$^x$ is formed.

The circuit that computes the power terms is shown in fig. 4.2. There are three registers: in reg. 2 we store $|x\rangle = |x_1 x_2 ....... x_n\rangle$, the state of a system of n qubits where is encoded the value of the number of which we want to compute the k-th order power and $|x\rangle$ is a vector of the computational basis in the Hilbert space generated by n qubits. Reg. 1 and 3 both contain $|0\rangle$, the fundamental state. In a first step, the state in reg. 2 is copied in reg. 1 by using a number of C-NOT gates (for simplicity we draw only one C-NOT; in fig. 4.3 a simple example is shown). By this way, using the next gate, it is possible to realize the multiplication and obtain $x^2$, stored in $|x^2\rangle$. After the transformation, in reg. 1 there is the fundamental state $|0\rangle$ while reg. 2 is left unchanged; it is obvious that reg. 3 should contain a sufficient number of qubits for storing $|x^2\rangle$. Next step consists in calculating $x^3$; a way for saving memory consists in using again reg. 1 by a "snake" technique; we consider the multiplication circuit and adapt it changing the structure of memory registers (see fig 4.4). The arrow from left to right indicates that the accumulator register is R1 and that R3 is in the fundamental state at the end of the multiplication procedure (no longer R1); in addition, there are two SWAP gates for every register: one is needed in the input to invert the order of the qubits and obtain the correct result of the calculation (this is due to the structure of the gate); the other (for the output of the gate) is necessary because qubits find in the next step an ordinary multiplication module (which gives $x^4$). So, power series terms are stored alternatively in R1 or R3; in the former we find odd terms, in the latter even terms. There are evident advantages: we only need three registers to execute the calculation; moreover, there is no need to put to zero artificially the intermediate registers because it is the procedure itself to make it for us: so the appropriate memory location is "formatted" for next steps.

In the following we will adopt a convention: each step of the multiplication is indicated as in fig. 4.5. This way, two different symbols are reduced to one; the symbol $|x^l\rangle$ denotes the output state of multiplication gate and it can represent both even and odd power terms;

one only needs to be careful about the register where it is stored (R1 for even power, R2 for odd).

At last power terms have to be "weighted" by 1/k! factors. It is clear that for $k>1$ 1/k! is less than 1. If we use binary fractions it is possible to approximate this number with arbitrary precision (it is sufficient to increase the number of qubits). We see some examples that will be useful in the following (binary fractions are indicated by a subscript 2 to distinguish them from decimal fractions):

1) $k = 2 \rightarrow \dfrac{1}{2!} = \dfrac{1}{2}$

It is obvious that:

$$\dfrac{1}{2!} = 0.1_2 \qquad (4.2)$$

2) $k = 3 \rightarrow \dfrac{1}{3!} = \dfrac{1}{6} = 0.1666\overline{6}$

Using eleven qubits we have:

$$\dfrac{1}{3!} \approx 0.00101010101_2 \approx 0.166504 \qquad (4.3)$$

The error in its evaluation is about 1‰.

If we consider bigger values for k, the number of qubits needed for the approximation must be increased:

3) $k = 6 \rightarrow \dfrac{1}{6!} = \dfrac{1}{720} = 0.001388\overline{8}$

Using 15 qubits:

$$\dfrac{1}{6!} \approx 0.000000000101101_2 \approx 0.0013734 \qquad (4.4)$$

whose evaluation error is about 1%.

A useful way for representing fractionary numbers is "floating point representation". By this technique it is possible to decrease the memory necessary to execute the calculations and simplify the circuits for arithmetic operations. In general, numbers are represented in the following way:

$$N = m \times b^e \tag{4.5}$$

where N in the number we want to write, $m$ is said "mantissa", $e$ is the "exponent" and $b$ the "base"; changing the value of $e$ it is possible to "move" the point. For numbers in digital representation the base is obviously 2; for instance:

$$1.5 = 1.1_2 = 11 \times 2^{-1}{}_2$$

In a computer, mantissa and exponent are stored in distinct memory sectors (whose sizes depend on the value of the numbers used and the precision of the calculation). A product between two binary numbers in floating point representation has this form:

$$N_1 = m_1 \times 2^{e_1}, \ N_2 = m_2 \times 2^{e_2} \rightarrow N_1 N_2 = (m_1 \times m_2) \times 2^{e_1 + e_2} \tag{4.6a}$$

For summation we need that the exponents of the numbers are the same. More explicitly:

$$N_1' = m_1' \times 2^{e_1}, \ N_2' = m_2' \times 2^{e_2} = m_2' \times 2^{e_1 + e_3} = m_2' \times 2^{e_3} \times 2^{e_1} = m_2'' \times 2^{e_1} \rightarrow$$

$$\rightarrow N_1' + N_2' = (m_1' + m_2'') \times 2^{e_1} \tag{4.7b}$$

We can now analyze the circuit in fig. 4.6.

We consider four registers. In R1 e R2 we store $|0\rangle = |00....0\rangle$ (i.e. a sufficient number of qubits for our purpose); R3 is composed by one qubit which is in the $|1\rangle$ state (in the computational basis representation it reads $\begin{pmatrix} 0 \\ 1 \end{pmatrix}$); in R4 we find a state $|y\rangle$: due to the fact that this circuit is a small part of a bigger circuit we suppose that in $|y\rangle$ there is the result of previous steps.

Due to the convention of fig. 4.5, we consider the state $|x^k\rangle$ obtained from the circuit in fig. 4.2 and, by a set of C-Not gates, we copy it in R1. In the following we will use a procedure called "Phase Estimation" [1]; this technique is based on qFT and makes possible to obtain an approximated value of φ from the eigenvalue $e^{i\varphi}$ of an unitary operator U; in our case, the phase φ is the number 1/k! codified as a binary fraction, the eigenstate $|u\rangle$ relative to $e^{i\varphi}$ is the state $|1\rangle$ and the operator U is the single-qubit gate

$$\begin{pmatrix} 1 & 0 \\ 0 & e^{2\pi i\varphi} \end{pmatrix} \tag{4.8}$$

The phase-estimation procedure stores in R2 the state $|\varphi_1\varphi_2...\varphi_t\rangle = |\varphi_k\rangle$ in which the parameter *t* (the number of qubits used to codify the phase) depends on the precision by which we have approximated 1/k!. We have to notice that, due to the structure of the extraction algorithm, the number stored in $|\varphi_k\rangle$ does not represent the binary fraction of 1/k! but the number $(0.\varphi_1\varphi_2....\varphi_t)_2 \times 2^t$. As shown in the case of numbers in floating point, it is necessary to consider a supplementary register (not shown, for simplicity, in fig 4.6) where the value of the exponent has to be memorized (in our case the number –*t*); its size depends on the precision of our approximation of the binary fraction. For instance, if we establish that all the weights of the power expansion have to be approximated with 15 qubits, we can store the number -15 and go on with the calculation keeping fixed the value of the register. Next step is the multiplication of numbers encoded in $|\varphi_k\rangle$ and $|x^k\rangle$ storing the result $|\varphi_k x^k\rangle$ in R4; the multiplication algorithm used here works by successive summations in the accumulator register R4; so, in our case, $|\varphi_k x^k\rangle$ is added to $|y\rangle$. We have to remember that summation of numbers in floating point can be done only if the exponents of the addends are the same; as a consequence, in R4 $|y\rangle$ is codified by a bigger number of qubits. An example will clarify this point: we suppose that $|y\rangle = |10\rangle = |1\rangle|0\rangle$ representing number 2; if we want to add to it the number $N = 3 \times 2^{-2}$, we need that it has the exponent -2; so we have $2 = 8 \times 2^{-2}$; now the mantissa is 8 and its representative state is $|y'\rangle = |1000\rangle = |1\rangle|0\rangle|0\rangle|0\rangle$; it is sufficient to add two less significant qubits in the register to make the summation possible without any problem; the register that formerly contained the exponent of number N now is used to store the *y* exponent too. If we set a fixed value for the exponent of $\varphi_k$, the number of $|0\rangle$ states to add to the register R4 is fixed too and we can neglect it in the following.

After the multiplication, in R1 there is the fundamental state $|0\rangle$ (ready for the storing of $|x^{k+1}\rangle$ in the next step); in R3 there is still the state $|1\rangle$ that is not modified by the phase estimation algorithm; in R4 there is the state $|y + \varphi_k x^k\rangle$, final result of the procedure; in R2 there is $|\varphi_k\rangle$; due to the fact that we want the fundamental state $|0\rangle = |00....0\rangle$ once again, it is necessary to put to zero $|\varphi_k\rangle$. The circuit in fig. 4.7 (which, for simplicity, is lacking of the initial qFT circuit) is able to accomplish the required procedure; its structure is symmetric to that of phase estimation [1]; the gates labelled $-U^{2^k}$ has a matrix representation similar to that of (4.8):

$$\begin{pmatrix} 1 & 0 \\ 0 & e^{-2\pi i \varphi} \end{pmatrix} \tag{4.9}$$

A an example, let us consider the state $|0\rangle + e^{2\pi i (2^k \varphi)}|1\rangle$; the action of $-U^{2^k}$ on its eigenstate $|1\rangle$ is:

$$(-U^{2^k})|1\rangle = \begin{pmatrix} 1 & 0 \\ 0 & e^{-2\pi i \varphi} \end{pmatrix}^{2^k} \begin{pmatrix} 0 \\ 1 \end{pmatrix} = e^{-2\pi i (2^k \varphi)} \begin{pmatrix} 0 \\ 1 \end{pmatrix} \tag{4.10a}$$

If we consider the behaviour of a controlled gate we have:

$$(C-(-U^{2^k}))[(|0\rangle + e^{2\pi i(2^k \varphi)}|1\rangle)|1\rangle] = (|0\rangle + e^{-2\pi i(2^k \varphi)}e^{2\pi i(2^k \varphi)}|1\rangle)|1\rangle =$$

$$= (|0\rangle + |1\rangle)|1\rangle \xrightarrow{H} |0\rangle|1\rangle \tag{4.10b}$$

In the end, this procedure permits to obtain the product of the power term and of its appropriate weight in the expansion (3.1) and then to sum it to the result of the previous steps; the advantage is given by the use of only four registers (neglecting the one needed to store the exponent), one of which is constituted by one qubit; at the end of each step, R1, R2 and R3 are again in the initial state, ready to be used in the next step.

**5 – An example of the approximation procedure**

As an example of the whole procedure, we execute the approximate calculation of the function $e^x$ with x = 2. Because of its length, we limit our calculation to the second order of the series expansion; it is obvious that such an approximation is brute and gives a result that is relatively far from the real value ($e^2 \approx 7.3891$). Our second order development gives a result of 5; if we go to the fourth order we obtain 7.

A further observation before of going on: in the following we suppose to have in every register a number of qubits that is sufficient for our purpose. Moreover, by convention, most significant qubits are omitted if they have 0 value and they are not necessary for a cogent execution of the procedure; for instance:

$$|1\rangle \equiv |01\rangle \equiv |001\rangle \equiv ... \equiv |00...01\rangle \tag{5.1}$$

It is clear that this writing is improper but it simplifies the calculations; in the case of ambiguity, we will show the correct states.

For purpose of clarity, we have divided the procedure in two distinct paragraphs. First of all we compute the value of the power terms needed in our expansion. Afterwards, we "weight" each term with the appropriate coefficient and obtain the approximated result.

**5.1 – Calculation of power terms**

We calculate $2^2$ starting from 2.

We store the state $|010\rangle$ (number 2) in the register of the multiplicand; in the multiplier and in the accumulator register we store the fundamental state $|000\rangle$; the control qubit is set to zero. The content of the multiplicand is copied in the multiplier and, after the check of the control register (it verifies if there is 0 in the multiplier register), the qFT is applied to the accumulator register:

$$|000\rangle = |0\rangle|0\rangle|0\rangle \xrightarrow{qFT} \begin{cases} \frac{1}{\sqrt{2}}(|0\rangle + e^{2\pi i(0/2+0/4+0/8)}|1\rangle) \\ \frac{1}{\sqrt{2}}(|0\rangle + e^{2\pi i(0/2+0/4)}|1\rangle) \\ \frac{1}{\sqrt{2}}(|0\rangle + e^{2\pi i(0/2)}|1\rangle) \end{cases} \quad (5.2)$$

Multiplicand is then added to the accumulator:

$$(5.2) \xrightarrow{SUM} \begin{cases} \frac{1}{\sqrt{2}}(|0\rangle + e^{2\pi i(0/2+0/4+0/8+0/2+1/4+0/8)}|1\rangle) \\ \frac{1}{\sqrt{2}}(|0\rangle + e^{2\pi i(0/2+0/4+1/2+0/4)}|1\rangle) \\ \frac{1}{\sqrt{2}}(|0\rangle + e^{2\pi i(0/2+0/2)}|1\rangle) \end{cases} = \begin{cases} \frac{1}{\sqrt{2}}(|0\rangle + e^{2\pi i(0/2+1/4+0/8)}|1\rangle) \\ \frac{1}{\sqrt{2}}(|0\rangle + e^{2\pi i(1/2+0/4)}|1\rangle) \\ \frac{1}{\sqrt{2}}(|0\rangle + e^{2\pi i(0/2)}|1\rangle) \end{cases} \quad (5.3)$$

The content of the multiplier is decreased by 1 with by the gate in fig. 4.3:

$$|010\rangle \xrightarrow{qFT} \begin{cases} \frac{1}{\sqrt{2}}(|0\rangle + e^{2\pi i(0/2+1/4+0/8)}|1\rangle) \\ \frac{1}{\sqrt{2}}(|0\rangle + e^{2\pi i(1/2+0/4)}|1\rangle) \\ \frac{1}{\sqrt{2}}(|0\rangle + e^{2\pi i(0/2)}|1\rangle) \end{cases} \xrightarrow{Decrement}$$

$$\xrightarrow{Decrement} \begin{cases} \frac{1}{\sqrt{2}}(|0\rangle + e^{2\pi i(0/2+1/4+0/8-1/8)}|1\rangle) \\ \frac{1}{\sqrt{2}}(|0\rangle + e^{2\pi i(1/2+0/4-1/4)}|1\rangle) \\ \frac{1}{\sqrt{2}}(|0\rangle + e^{2\pi i(0/2+1/2)}|1\rangle) \end{cases} = \begin{cases} \frac{1}{\sqrt{2}}(|0\rangle + e^{2\pi i(0/2+0/4+1/8)}|1\rangle) \\ \frac{1}{\sqrt{2}}(|0\rangle + e^{2\pi i(0/2+1/4)}|1\rangle) \\ \frac{1}{\sqrt{2}}(|0\rangle + e^{2\pi i(1/2)}|1\rangle) \end{cases}$$

$$\xrightarrow{qFT^{-1}} |001\rangle \quad (5.4)$$

Since the multiplier is not yet zero, the multiplication procedure requires another summation between multiplicand and accumulator:

$$(5.3) \xrightarrow{SUM} \begin{cases} \frac{1}{\sqrt{2}}(|0\rangle + e^{2\pi i(0/2 + 1/4 + 0/8 + 0/2 + 1/4 + 0/8)}|1\rangle) \\ \frac{1}{\sqrt{2}}(|0\rangle + e^{2\pi i(1/2 + 0/4 + 1/2 + 0/4)}|1\rangle) \\ \frac{1}{\sqrt{2}}(|0\rangle + e^{2\pi i(0/2 + 0/2)}|1\rangle) \end{cases} =$$

$$= \begin{cases} \frac{1}{\sqrt{2}}(|0\rangle + e^{2\pi i(1/2 + 0/4 + 0/8)}|1\rangle) \\ \frac{1}{\sqrt{2}}(|0\rangle + e^{2\pi i(1 + 0/2 + 0/4)}|1\rangle) \\ \frac{1}{\sqrt{2}}(|0\rangle + e^{2\pi i(0/2)}|1\rangle) \end{cases} = \begin{cases} \frac{1}{\sqrt{2}}(|0\rangle + e^{2\pi i(1/2 + 0/4 + 0/8)}|1\rangle) \\ \frac{1}{\sqrt{2}}(|0\rangle + e^{2\pi i(0/2 + 0/4)}|1\rangle) \\ \frac{1}{\sqrt{2}}(|0\rangle + e^{2\pi i(0/2)}|1\rangle) \end{cases} \quad (5.5)$$

There is another decrement operation on multiplier:

$$|001\rangle \xrightarrow{qFT} \begin{cases} \frac{1}{\sqrt{2}}(|0\rangle + e^{2\pi i(0/2 + 0/4 + 1/8)}|1\rangle) \\ \frac{1}{\sqrt{2}}(|0\rangle + e^{2\pi i(0/2 + 1/4)}|1\rangle) \\ \frac{1}{\sqrt{2}}(|0\rangle + e^{2\pi i(1/2)}|1\rangle) \end{cases} \xrightarrow{Decrement}$$

$$\xrightarrow{Decrement} \begin{cases} \frac{1}{\sqrt{2}}(|0\rangle + e^{2\pi i(0/2 + 0/4 + 1/8 - 1/8)}|1\rangle) \\ \frac{1}{\sqrt{2}}(|0\rangle + e^{2\pi i(0/2 + 1/4 - 1/4)}|1\rangle) \\ \frac{1}{\sqrt{2}}(|0\rangle + e^{2\pi i(1/2 + 1/2)}|1\rangle) \end{cases} = \begin{cases} \frac{1}{\sqrt{2}}(|0\rangle + e^{2\pi i(0/2 + 0/4 + 0/8)}|1\rangle) \\ \frac{1}{\sqrt{2}}(|0\rangle + e^{2\pi i(0/2 + 0/4)}|1\rangle) \\ \frac{1}{\sqrt{2}}(|0\rangle + e^{2\pi i(1 + 0/2)}|1\rangle) \end{cases}$$

$$\xrightarrow{qFT^{-1}} |000\rangle \quad (5.6)$$

If we check the multiplier, the control qubit passes from the state $|0\rangle$ to the state $|1\rangle$ because now the register is zero; this way, we inhibit another summation of multiplicand and accumulator, enabling on the latter the inverse qFT procedure; then we get:

$$(5.5) \xrightarrow{qFT^{-1}} |100\rangle \quad (5.7)$$

Implementation of analytic functions with quantum gates

The state (5.7) represents the decimal number 4 (in binary notation): then, the circuit is able to compute the power terms. If we further apply the procedure (now using the circuit in fig. 4.4) we can compute the value $2^3 = 8$.

**5.2 – "Weighted" power terms and the calculation of the approximated function**

It is obvious that power terms of order 0 and 1 are multiplied by "weights" 1. They can be immediately added in R4 of fig. 4.6; in our case we have:

1) $x^0 = 2^0 = 1$ that is codified in the state $|01\rangle$ (we neglect most significant qubits in the state $|0\rangle$).

2) $x^1 = 2^1 = 2$ that is codified in the state $|10\rangle$.

We suppose that initially there is the state $|01\rangle$ in R4; we can sum $|10\rangle$ to it by the insertion of an summation circuit between the second register of the power terms module (where the number 2 is always stored) and R4; explicitly we have:

$$|01\rangle \xrightarrow{qFT} \begin{cases} \frac{1}{\sqrt{2}}(|0\rangle + e^{2\pi i(0/2 + 1/4)}|1\rangle) \\ \frac{1}{\sqrt{2}}(|0\rangle + e^{2\pi i(1/2)}|1\rangle) \end{cases} \xrightarrow{SUM} \begin{cases} \frac{1}{\sqrt{2}}(|0\rangle + e^{2\pi i(0/2 + 1/4 + 1/2 + 0/4)}|1\rangle) \\ \frac{1}{\sqrt{2}}(|0\rangle + e^{2\pi i(1/2 + 0/2)}|1\rangle) \end{cases} =$$

$$= \begin{cases} \frac{1}{\sqrt{2}}(|0\rangle + e^{2\pi i(1/2 + 1/4)}|1\rangle) \\ \frac{1}{\sqrt{2}}(|0\rangle + e^{2\pi i(1/2)}|1\rangle) \end{cases} \xrightarrow{qFT^{-1}} |11\rangle \qquad (5.8)$$

The state (5.8) represents the number 3 that we expected.

The power term of order 2 must be multiplied by 1/2; in this case the approximation of the weight by a binary fraction is exact because we have:

$$\frac{1}{2} = 0.1_2 = 2^{-1} \qquad (5.9)$$

The state where is codified number 4 is copied in R1 of fig. 4.6. the U operator where is stored the binary fraction takes the form:

$$\begin{pmatrix} 1 & 0 \\ 0 & e^{2\pi i 0.1} \end{pmatrix} = \begin{pmatrix} 1 & 0 \\ 0 & e^{2\pi i (\frac{1}{2})} \end{pmatrix} \qquad (5.10)$$

Due to the fact that one qubit only is needed to codify 1/2, the circuit for phase estimation is composed by two qubits: one for the eigenstate $|1\rangle$ and the other for the auxiliary register; explicitly:

$$|0\rangle|1\rangle \xrightarrow{H_1} \frac{1}{\sqrt{2}}(|0\rangle + |1\rangle)|1\rangle \xrightarrow{C-U} \frac{1}{\sqrt{2}}(|0\rangle + e^{2\pi i (\frac{1}{2})}|1\rangle)|1\rangle \xrightarrow{(qFT^{-1})_1}$$

$$\xrightarrow{(qFT^{-1})_1} |1\rangle|1\rangle \qquad (5.11)$$

At first sight, this result looks trivial; however, if we think of more complicated situations where the phase approximation requires the use of a big number of qubits, we see immediately that this procedure is less trivial if executed by hand.

As explained in the general discussion about the algorithm, binary fractions are stored in a state; in our case it represents number 1; for obtaining the expected result it is necessary to arrange a register for storing the exponent of the floating point representation (-1 in our case); for our purpose two qubits are enough: one for the sign (by convention and in analogy with the classical case, we set $|1\rangle$ for the negative sign) and the other to memorize the absolute value of the exponent; this way, the number is represented as:

$$1 \times 4 \times 2^{-1} = 2 \qquad (5.12)$$

We must remember that in R4 there is number 3 coded in state $|11\rangle$; to get floating point sum to be correct, number 3 must be written as

Implementation of analytic functions with quantum gates

$$3 = 3 \times 2 \times 2^{-1} = 6 \times 2^{-1} \tag{5.13}$$

Thus, we store number 6 in R1 by the state $|110\rangle$; in order to obtain this state, we can add in R1 another state $|0\rangle$ as least significant qubit (we could consider a qubit we have neglected until now).

Let us consider now the sum and multiplication procedure. Let us suppose that R1 is the multiplier and R2 is the multiplicand; qFT is applied to R4 and we have (considering 0 as most significant qubit thanks to our convention):

$$|0110\rangle \xrightarrow{QFT} \begin{cases} \frac{1}{\sqrt{2}}(|0\rangle + e^{2\pi i(0/2 + 1/4 + 1/8 + 0/16)}|1\rangle) \\ \frac{1}{\sqrt{2}}(|0\rangle + e^{2\pi i(1/2 + 1/4 + 0/8)}|1\rangle) \\ \frac{1}{\sqrt{2}}(|0\rangle + e^{2\pi i(1/2 + 0/4)}|1\rangle) \\ \frac{1}{\sqrt{2}}(|0\rangle + e^{2\pi i(0/2)}|1\rangle) \end{cases} \tag{5.14}$$

The check of the control qubit on R1 reveals a state $|0100\rangle$ and enables the sum of multiplicand ($|0001\rangle$) and accumulator (R4); we have:

$$(5.14) \xrightarrow{SUM} \begin{cases} \frac{1}{\sqrt{2}}(|0\rangle + e^{2\pi i(0/2 + 1/4 + 1/8 + 0/16 + 0/2 + 0/4 + 0/8 + 1/16)}|1\rangle) \\ \frac{1}{\sqrt{2}}(|0\rangle + e^{2\pi i(1/2 + 1/4 + 0/8 + 0/2 + 0/4 + 1/8)}|1\rangle) \\ \frac{1}{\sqrt{2}}(|0\rangle + e^{2\pi i(1/2 + 0/4 + 0/2 + 1/4)}|1\rangle) \\ \frac{1}{\sqrt{2}}(|0\rangle + e^{2\pi i(0/2 + 1/2)}|1\rangle) \end{cases} =$$

$$= \begin{cases} \frac{1}{\sqrt{2}}(|0\rangle + e^{2\pi i(0/2 + 1/4 + 1/8 + 1/16)}|1\rangle) \\ \frac{1}{\sqrt{2}}(|0\rangle + e^{2\pi i(1/2 + 1/4 + 1/8)}|1\rangle) \\ \frac{1}{\sqrt{2}}(|0\rangle + e^{2\pi i(1/2 + 1/4)}|1\rangle) \\ \frac{1}{\sqrt{2}}(|0\rangle + e^{2\pi i(1/2)}|1\rangle) \end{cases} \tag{5.15}$$

The multiplier is decreased by 1:

$$|0100\rangle \xrightarrow{qFT} \begin{cases} \frac{1}{\sqrt{2}}(|0\rangle + e^{2\pi i(0/2+1/4+0/8+0/16)}|1\rangle) \\ \frac{1}{\sqrt{2}}(|0\rangle + e^{2\pi i(1/2+0/4+0/8)}|1\rangle) \\ \frac{1}{\sqrt{2}}(|0\rangle + e^{2\pi i(0/2+0/4)}|1\rangle) \\ \frac{1}{\sqrt{2}}(|0\rangle + e^{2\pi i(0/2)}|1\rangle) \end{cases} \xrightarrow{Decrement}$$

$$\xrightarrow{Decrement} \begin{cases} \frac{1}{\sqrt{2}}(|0\rangle + e^{2\pi i(0/2+1/4+0/8+0/16-1/16)}|1\rangle) \\ \frac{1}{\sqrt{2}}(|0\rangle + e^{2\pi i(1/2+0/4+0/8-1/8)}|1\rangle) \\ \frac{1}{\sqrt{2}}(|0\rangle + e^{2\pi i(0/2+0/4-1/4)}|1\rangle) \\ \frac{1}{\sqrt{2}}(|0\rangle + e^{2\pi i(0/2+1/2)}|1\rangle) \end{cases} = \begin{cases} \frac{1}{\sqrt{2}}(|0\rangle + e^{2\pi i(0/2+0/4+1/8+1/16)}|1\rangle) \\ \frac{1}{\sqrt{2}}(|0\rangle + e^{2\pi i(0/2+1/4+1/8)}|1\rangle) \\ \frac{1}{\sqrt{2}}(|0\rangle + e^{2\pi i(0/2+1/4+1/8)}|1\rangle) \\ \frac{1}{\sqrt{2}}(|0\rangle + e^{2\pi i(1/2)}|1\rangle) \end{cases}$$

$$\xrightarrow{qFT^{-1}} |0011\rangle \qquad (5.16)$$

We can repeat the sum operation between multiplicand and accumulator because multiplier is not zero yet:

$$(5.15) \xrightarrow{SUM} \begin{cases} \frac{1}{\sqrt{2}}(|0\rangle + e^{2\pi i(0/2+1/4+1/8+1/16+0/2+0/4+0/8+1/16)}|1\rangle) \\ \frac{1}{\sqrt{2}}(|0\rangle + e^{2\pi i(1/2+1/4+1/8+0/2+0/4+1/8)}|1\rangle) \\ \frac{1}{\sqrt{2}}(|0\rangle + e^{2\pi i(1/2+1/4+0/2+1/4)}|1\rangle) \\ \frac{1}{\sqrt{2}}(|0\rangle + e^{2\pi i(1/2+1/2)}|1\rangle) \end{cases} =$$

Implementation of analytic functions with quantum gates

$$
= \begin{cases}
\frac{1}{\sqrt{2}}(|0\rangle + e^{2\pi i(1/2 + 0/4 + 0/8 + 0/16)}|1\rangle) \\
\frac{1}{\sqrt{2}}(|0\rangle + e^{2\pi i(0/2 + 0/4 + 0/8)}|1\rangle) \\
\frac{1}{\sqrt{2}}(|0\rangle + e^{2\pi i(0/2 + 0/4)}|1\rangle) \\
\frac{1}{\sqrt{2}}(|0\rangle + e^{2\pi i(0/2)}|1\rangle)
\end{cases}
\tag{5.17}
$$

The decrement operation on R1 transforms (5.16) into state

$$|0010\rangle \tag{5.18}$$

We repeat the sum:

$$
(5.17) \xrightarrow{SUM} \begin{cases}
\frac{1}{\sqrt{2}}(|0\rangle + e^{2\pi i(1/2 + 0/4 + 0/8 + 0/16 + 0/2 + 0/4 + 0/8 + 1/16)}|1\rangle) \\
\frac{1}{\sqrt{2}}(|0\rangle + e^{2\pi i(0/2 + 0/4 + 0/8 + 0/2 + 0/4 + 1/8)}|1\rangle) \\
\frac{1}{\sqrt{2}}(|0\rangle + e^{2\pi i(0/2 + 0/4 + 0/2 + 1/4)}|1\rangle) \\
\frac{1}{\sqrt{2}}(|0\rangle + e^{2\pi i(0/2 + 1/2)}|1\rangle)
\end{cases} =
$$

$$
= \begin{cases}
\frac{1}{\sqrt{2}}(|0\rangle + e^{2\pi i(1/2 + 0/4 + 0/8 + 1/16)}|1\rangle) \\
\frac{1}{\sqrt{2}}(|0\rangle + e^{2\pi i(0/2 + 0/4 + 1/8)}|1\rangle) \\
\frac{1}{\sqrt{2}}(|0\rangle + e^{2\pi i(0/2 + 1/4)}|1\rangle) \\
\frac{1}{\sqrt{2}}(|0\rangle + e^{2\pi i(1/2)}|1\rangle)
\end{cases}
\tag{5.19}
$$

Thus:

$$(5.18) \xrightarrow{Decrement} |0001\rangle \tag{5.20}$$

Sum gives:

$$(5.19) \xrightarrow{SUM} \begin{cases} \frac{1}{\sqrt{2}}(|0\rangle + e^{2\pi i(1/2+0/4+0/8+1/16+0/2+0/4+0/8+1/16)}|1\rangle) \\ \frac{1}{\sqrt{2}}(|0\rangle + e^{2\pi i(0/2+0/4+1/8+0/2+0/4+1/8)}|1\rangle) \\ \frac{1}{\sqrt{2}}(|0\rangle + e^{2\pi i(0/2+1/4+0/2+1/4)}|1\rangle) \\ \frac{1}{\sqrt{2}}(|0\rangle + e^{2\pi i(1/2+1/2)}|1\rangle) \end{cases} =$$

$$= \begin{cases} \frac{1}{\sqrt{2}}(|0\rangle + e^{2\pi i(1/2+0/4+1/8+0/16)}|1\rangle) \\ \frac{1}{\sqrt{2}}(|0\rangle + e^{2\pi i(0/2+1/4+0/8)}|1\rangle) \\ \frac{1}{\sqrt{2}}(|0\rangle + e^{2\pi i(1/2+0/4)}|1\rangle) \\ \frac{1}{\sqrt{2}}(|0\rangle + e^{2\pi i(0/2)}|1\rangle) \end{cases} \qquad (5.21)$$

The decrement of the multiplier sets R1 in state 0.

Because of the zero setting, control qubit passes from $|0\rangle$ to $|1\rangle$ and enables the application of qFT$^{-1}$ on R4; in conclusion we have:

$$(5.21) \xrightarrow{qFT^{-1}} |1010\rangle \qquad (5.22)$$

By the "phase erasure" procedure (whose we gave an example in (2.10b)) it is possible to set R2 in the fundamental state.

In state (3.22) number 10 is coded; if we take into account the value of the exponent (-1, coded in another register), we see that the circuit has computed the number:

$$10 \times 2^{-1} = 5 \qquad (5.23)$$

5 is exactly the expected value for the 2nd order approximation of $e^2$. Successive orders can be obtained in the same way. Obviously they will require more calculations and more qubits in the registers. Notwithstanding this fact, they are all calculable, at least in principle, and give a result whose precision is comparable to that of a classical computer (in fact, the procedure is very similar in both cases). What is different is the core of the calculation sys-tem: as we have seen, a quantum computer can work with superposition of states and phases that can be added or erased in an easy way, leaving free memory in the register for executing other procedures, all without sacrifice of precision.

**6 – Conclusion**

In this paper we have shown a procedure to manipulate qubits so that they can be used to execute complex calculations and evaluate approximated values of an arbitrary analytic function.

In principle there are more efficient ways for the theoretical implementation of this recipe (for instance the multiplication step in less efficient than other known algorithms). Actually we have focused on simplicity for possible experimental implementations: repeated gates for simple operations can be realized with less efforts than complicated procedures.

We stress the fact that our analysis is limited to the procedure for executing the approximated calculation of a continuous and derivable function, neglecting physical problems such as decoherence and control during the evolution of the quantum system.

Encoding some information (numbers in our case but we can think to other applications) by vectors in a Hilbert space allows to use some particular properties and transformations. For instance, the application of a phase to a state is an operation that has no classical analogue; it has been seen that it is simpler adding and subtracting phases than copying the carries from a register to another. In the case of the circuit for the approximation, we have an evident advantage: it is possible, by only two registers, to execute the procedure of "power terms weighting", simply "unloading" each time the content of the memory by the execution of the "phase erasure" procedure.

In conclusion, the system shown in this paper could be taken as example of a ROM. The versatility of our circuit is ensured by the great variety of possible controlled-phase gates (changing the phase, we can change the weights and so the function we want to approximate).

**References**

1. M.I. Nielsen, I.A. Chuang, *Quantum computation and quantum information*, Cambridge University Press (2000).

2. P.W. Shor, *Algorithms for quantum computation: discrete logarithms and factoring; in Proceedings of the 35$^{th}$ Annual Symposium on the Foundations of Computer Science*, IEEE Computer Society Press, Los Alamitos, CA (1994).

3. L. Grover, *Proceedings of 28$^{th}$ Annual ACM Symposium on the Theory of Computation*, 212-219, ACM Press, N.Y. (1996).

4. L. Grover, *Quantum mechanics helps in searching for a needle in a haystack*, Phys. Rev. Lett., 79(2): 325, quant-ph/9706033 (1997).

5. V. Vedral, A. Barenco, A. Ekert, *Quantum networks for elementary arithmetic operations*, Phys. Rev A 54, 147. quant-ph/9511018 v1 (1995).

6. T.G. Draper, *Addition on a quantum computer*, quant-ph/0008033 (2000).

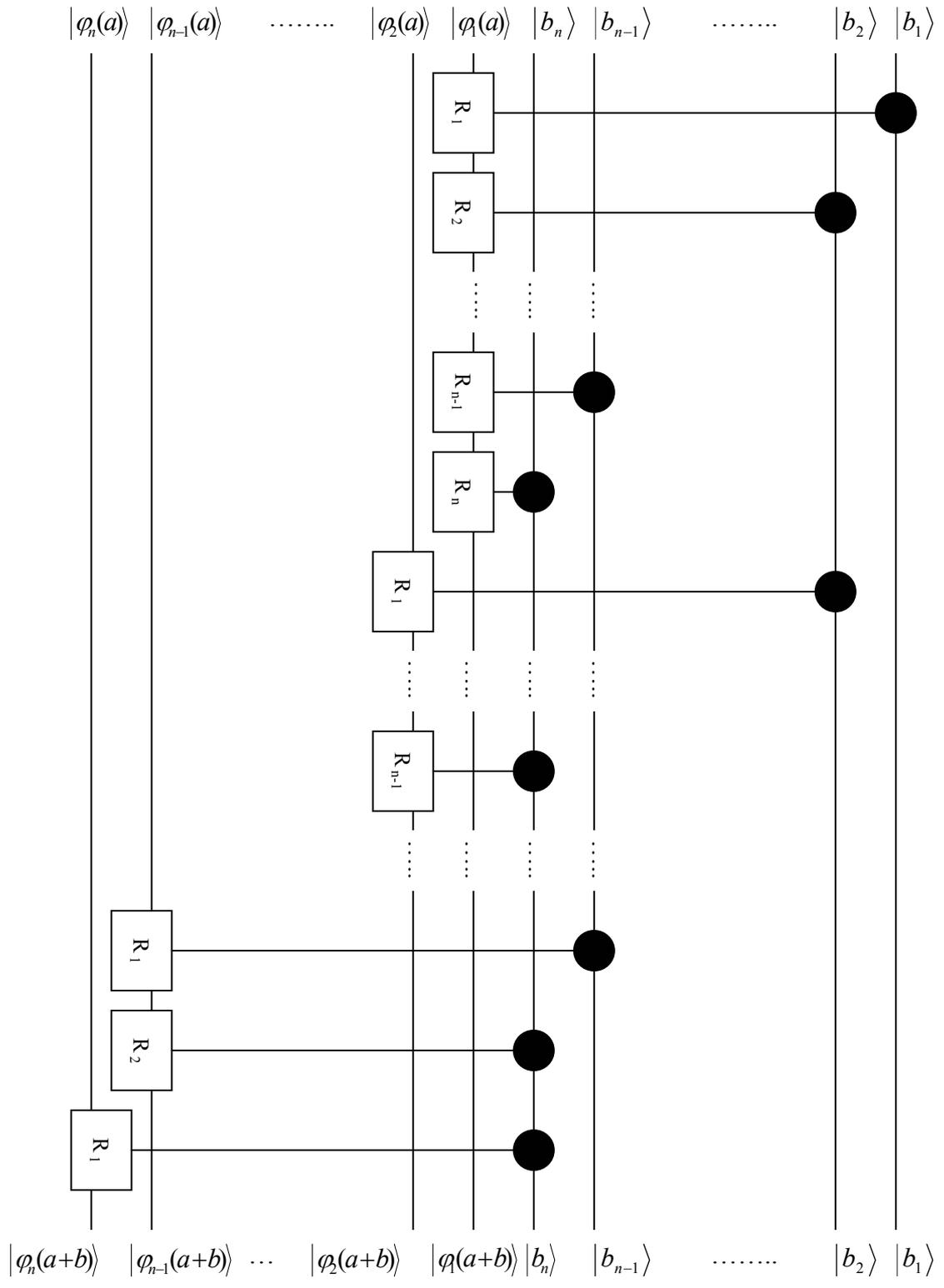

**Fig. 1.1** – Quantum Adder

Implementation of analytic functions with quantum gates

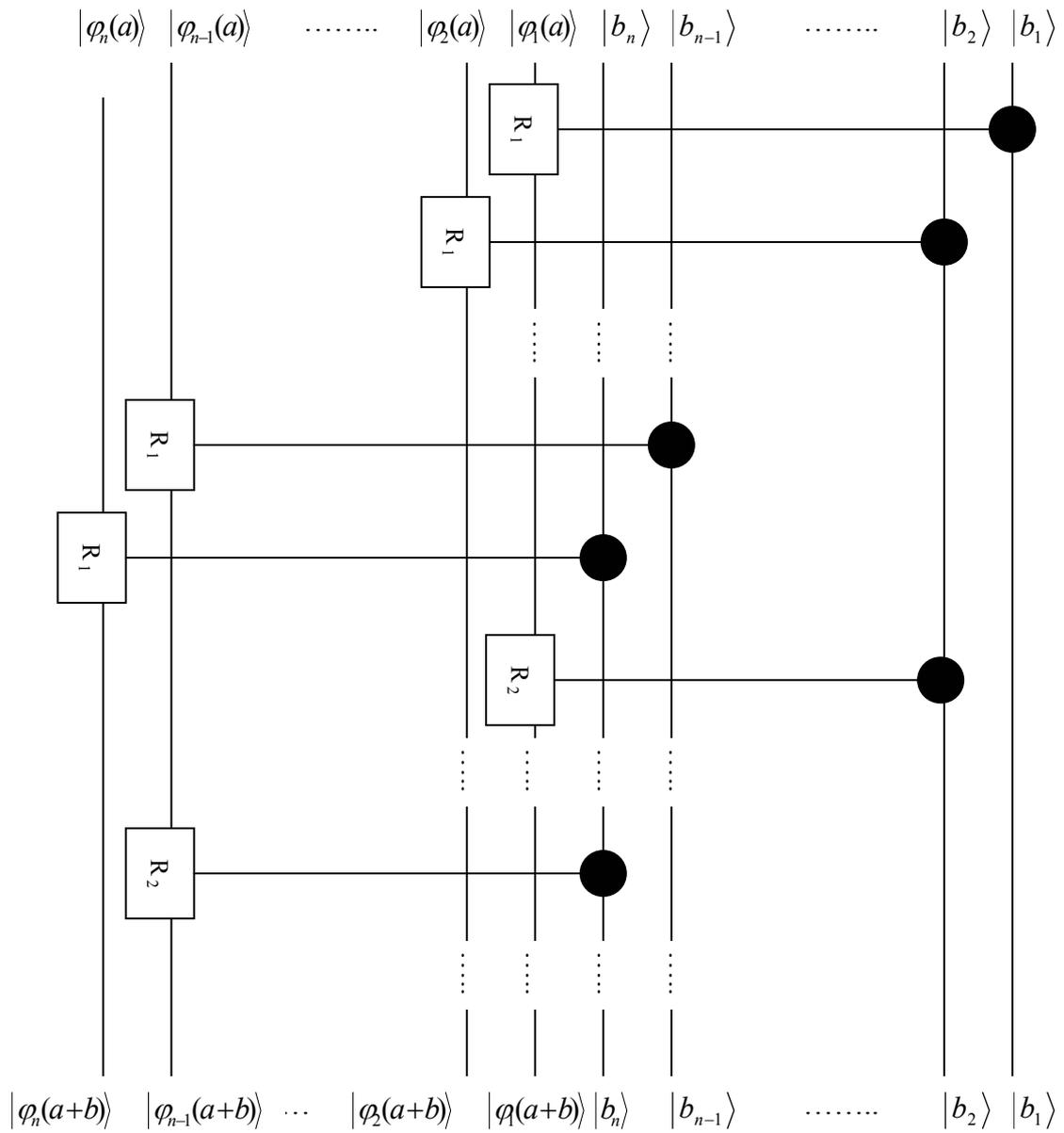

**Fig. 1.2** – Parallel Quantum Adder

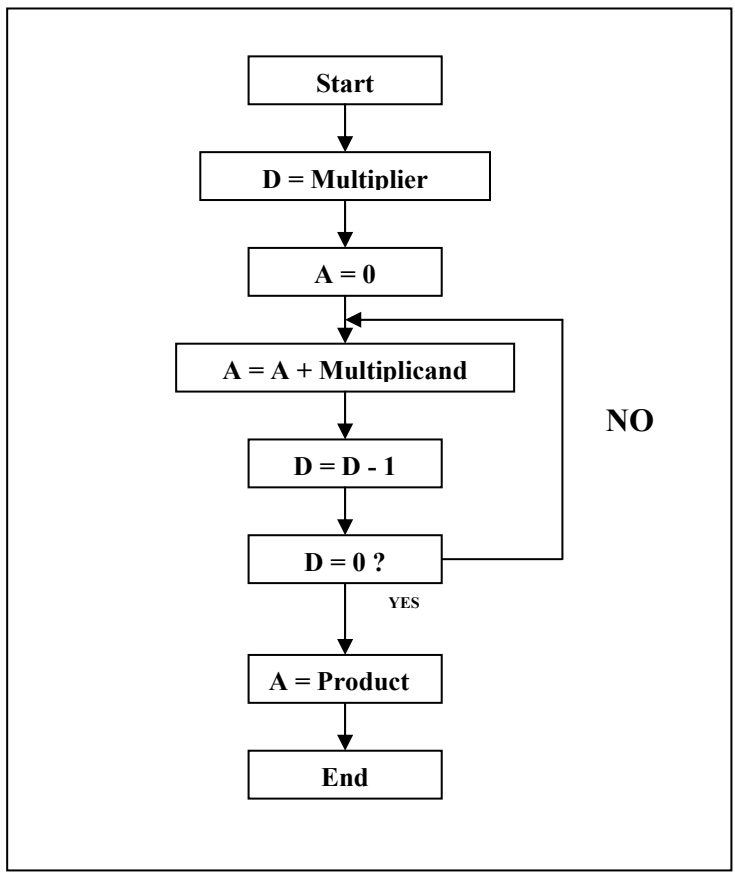

**Fig. 2.1** – Multiplication Flowchart

Implementation of analytic functions with quantum gates

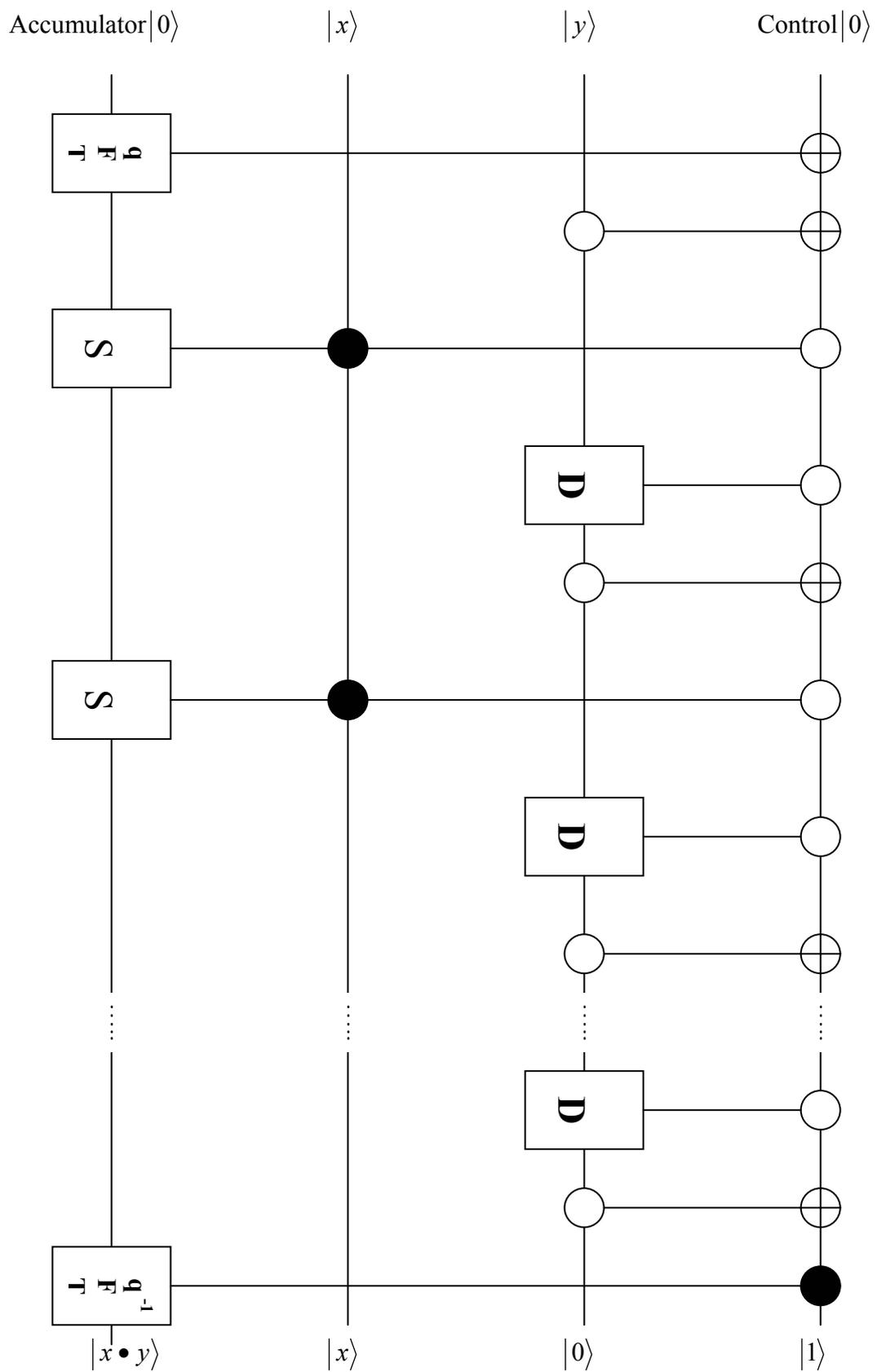

**Fig. 2.2** – Quantum Multiplier

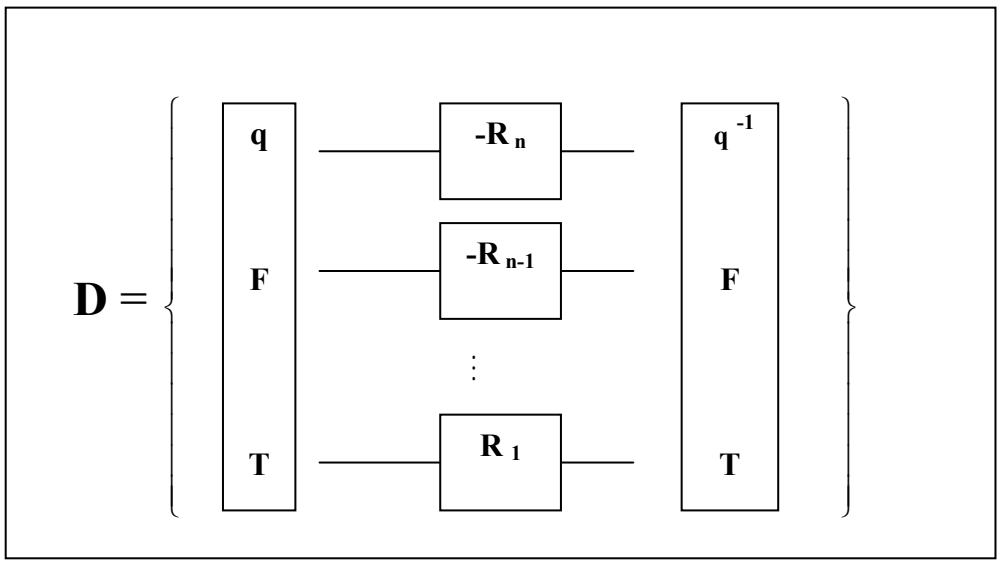

**Fig. 2.3** – Decrement Gate

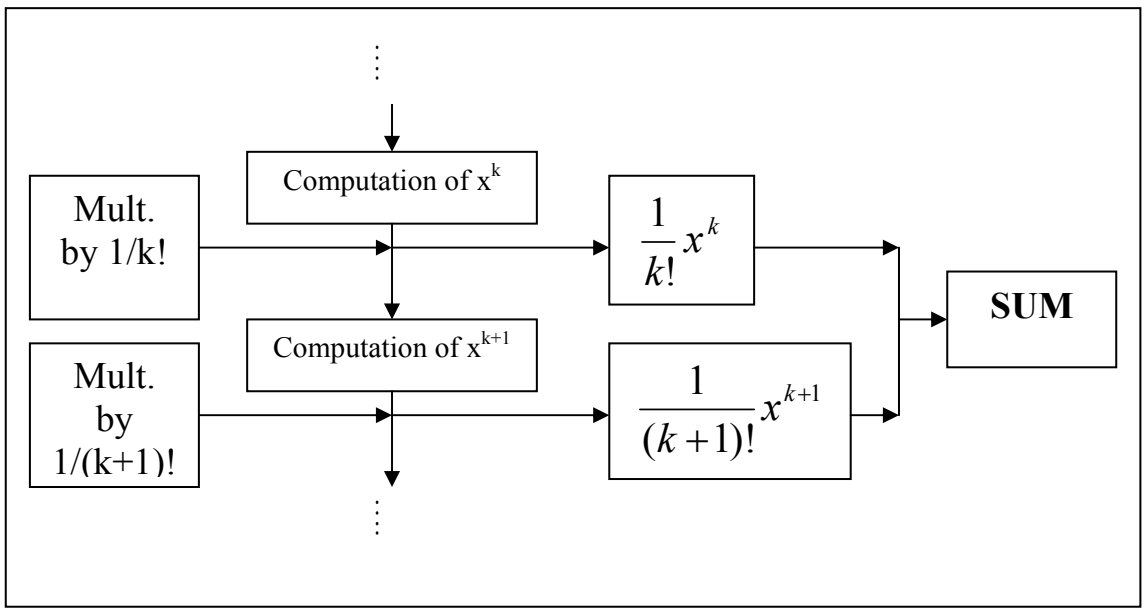

**Fig. 4.1** – Diagram for the evaluation of the function $e^x$

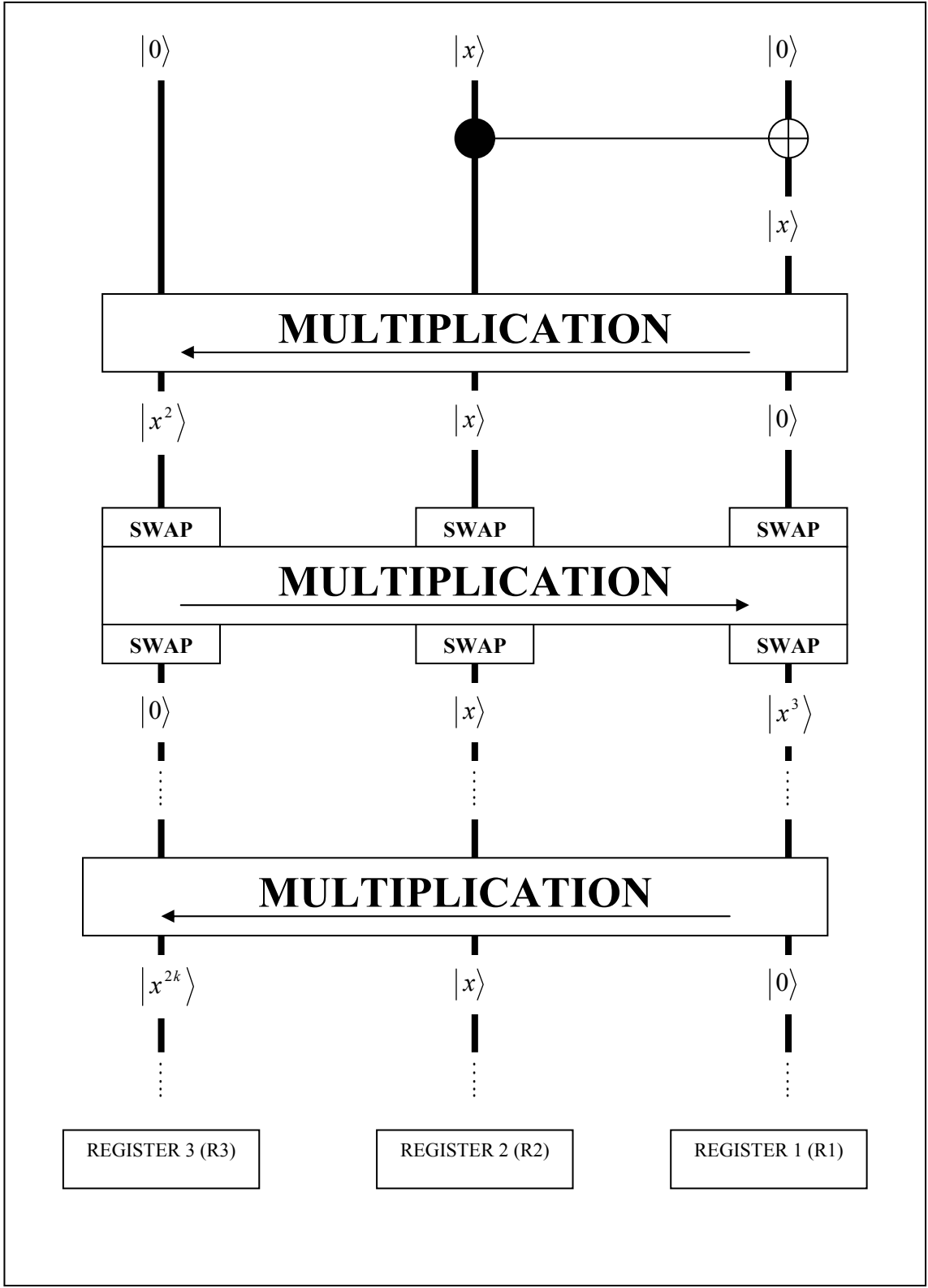

**Fig. 4.2** – "Snake" circuit for the calculation of power series terms

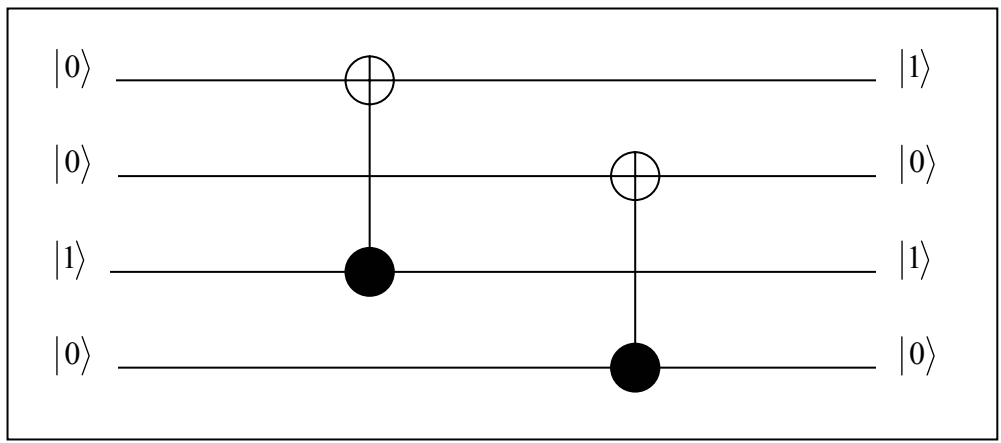

**Fig. 4.3** – C-NOT gates used to copy a memory register

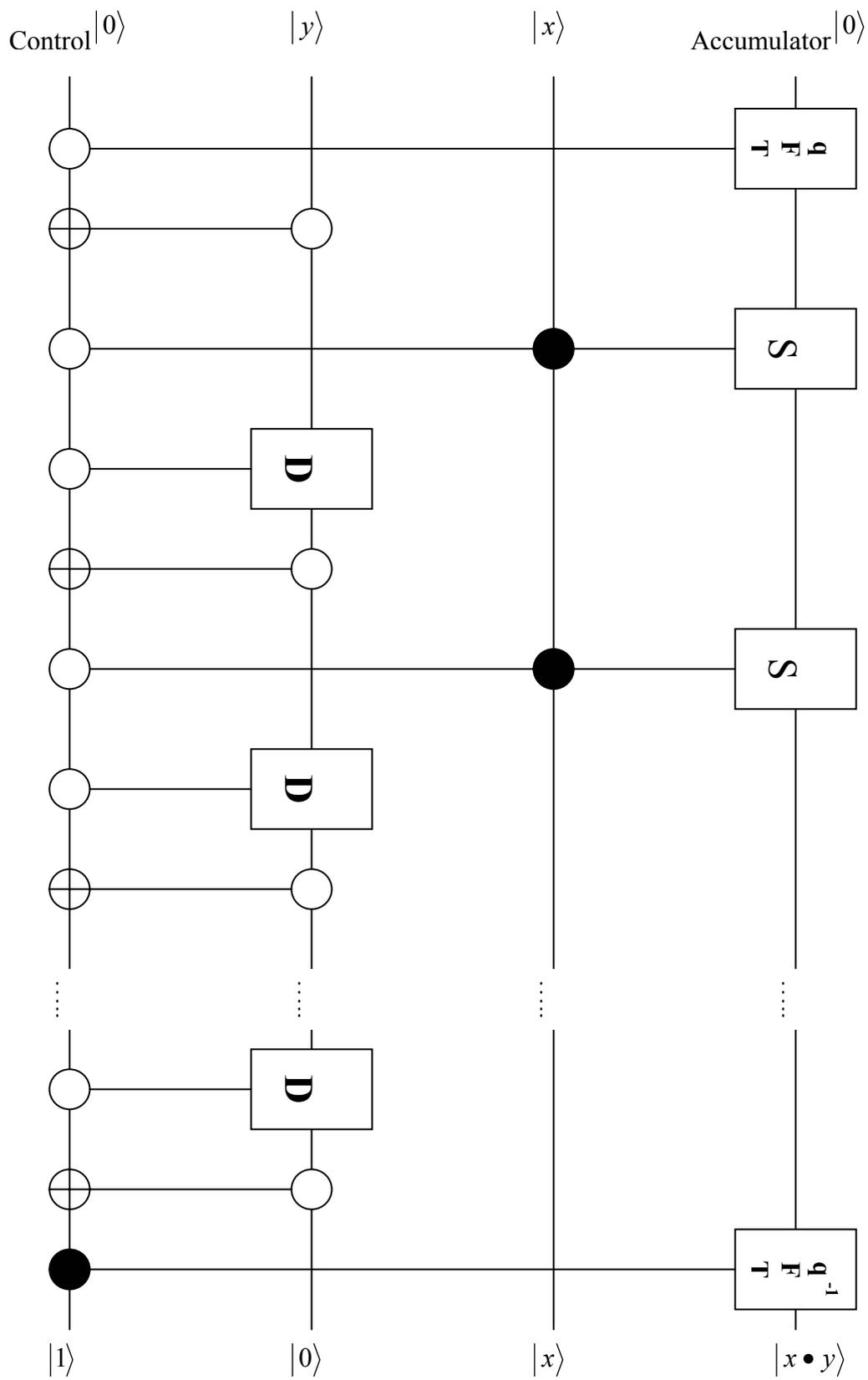

**Fig. 4.4** – Reversed circuit for quantum multiplication

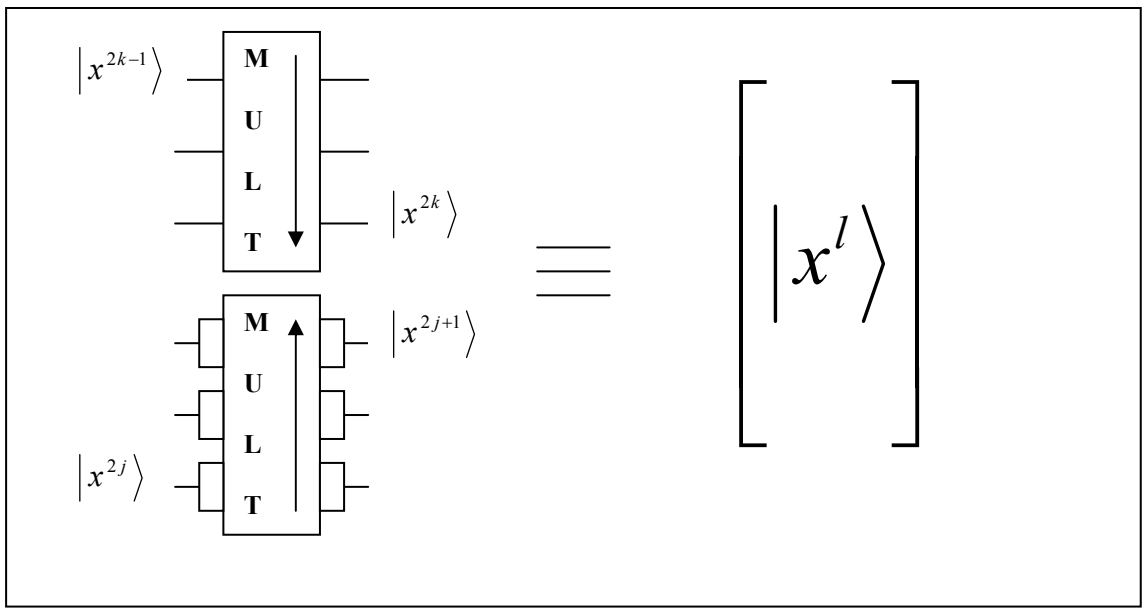

**Fig. 4.5** – Simplified representation of the multiplication gate

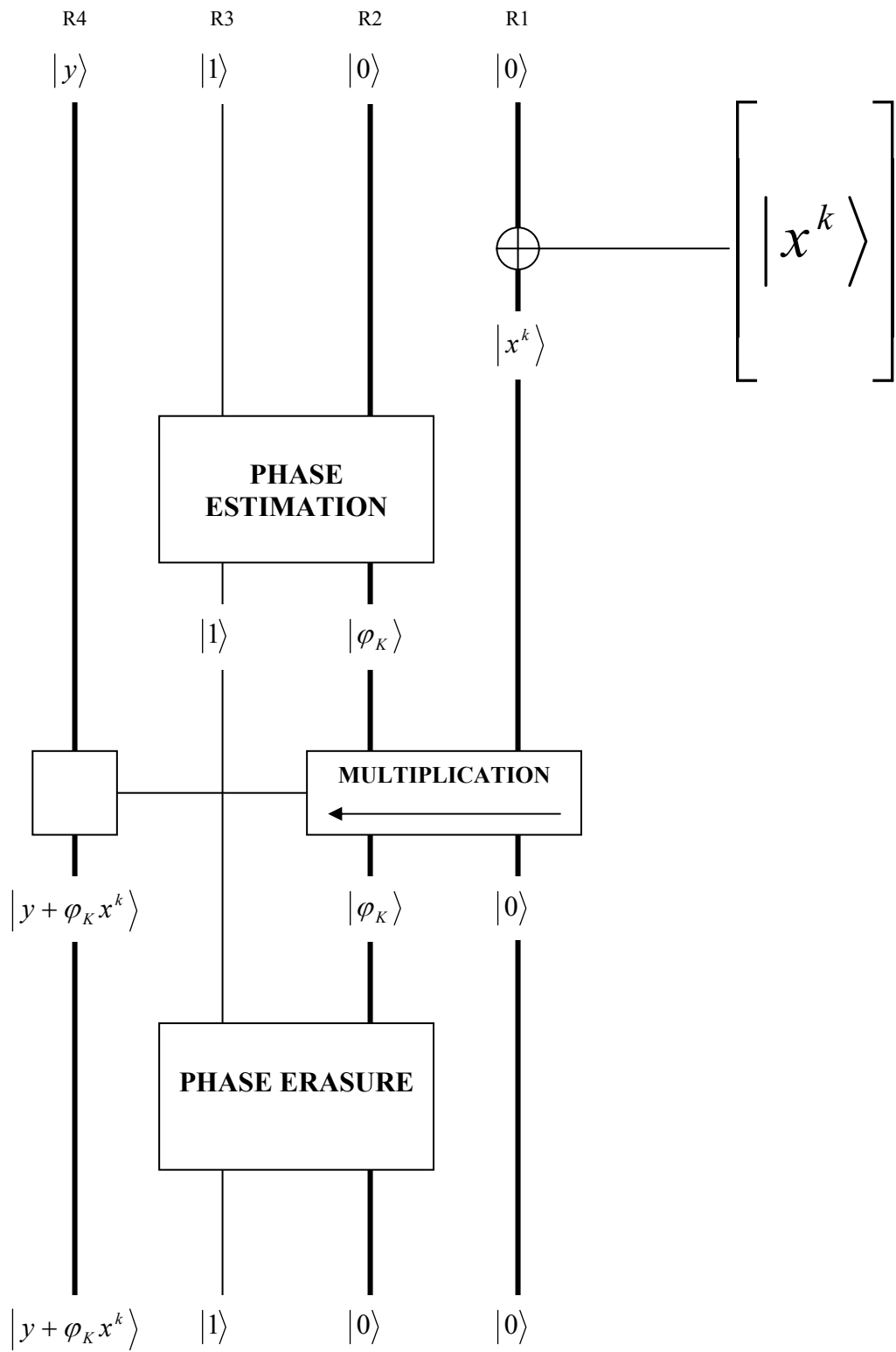

**Fig. 4.6** – circuit for the multiplication of weights 1/k!

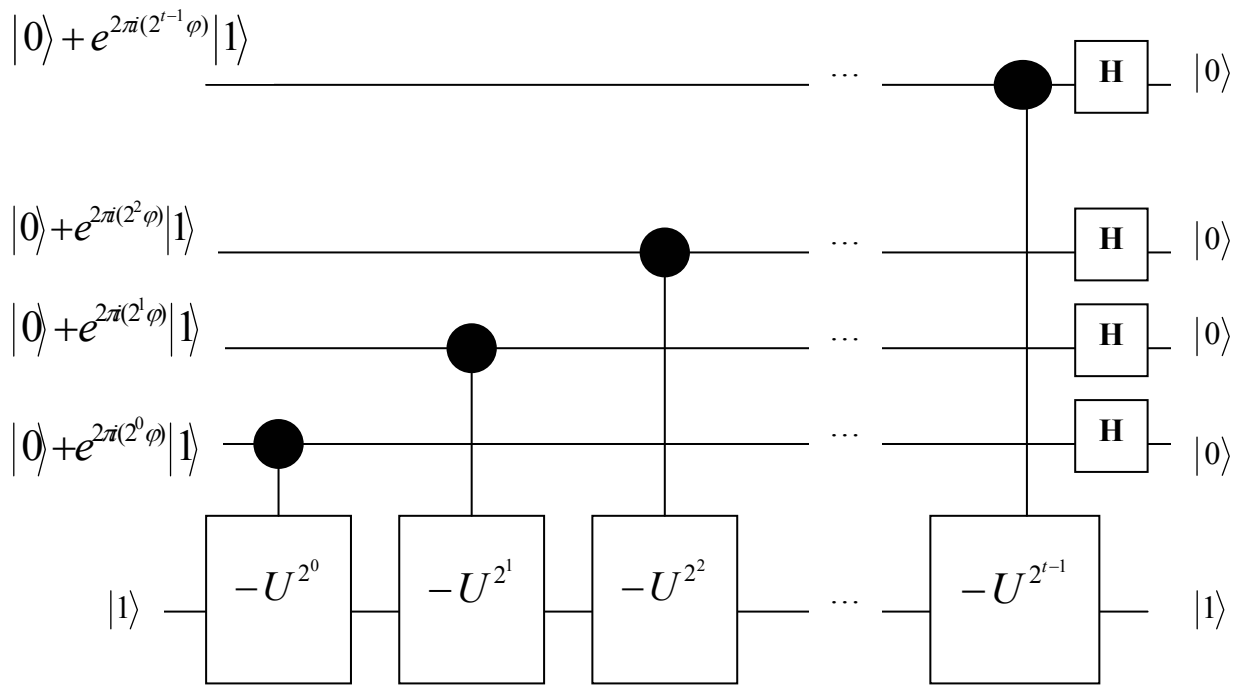

**Fig.4.7** – circuit for setting to zero $|\varphi_k\rangle$